\documentclass{aa}  

\usepackage{graphicx}
\usepackage{txfonts}
\usepackage{lscape}

\usepackage[colorlinks=true,citecolor=blue,linkcolor=blue]{hyperref}
\begin{document}

   \title{%
   Stellar pulsations interfering with the transit light curve: configurations with false positive misalignment
   }

   \author{
          A. B\'okon\inst{1}\fnmsep\thanks{\email{andrasb@titan.physx.u-szeged.hu}}
          \fnmsep\inst{2}\fnmsep\inst{3}
          \and
          Sz. K\'alm\'an\inst{1,4,5,6}
          \and
          I. B. B\'ir\'o\inst{7}\fnmsep\inst{8}
          \and
          M. Gy. Szab\'o\inst{1}\fnmsep\inst{9}
          }

   \institute{MTA-ELTE Exoplanet Research Group, Szent Imre h. u. 112, H-9700 Szombathely, Hungary
         \and
             Institute of Physics, Faculty of Sciences and Informatics, University of Szeged, D\'om t\'er 9, H-6720 Szeged, Hungary
         \and
             MTA-ELTE Lend\"ulet "Momentum" Milky Way Research Group, Hungary
         \and
             Konkoly Observatory, Research Centre for Astronomy and Earth Sciences, E\"otv\"os Lor\'and Research Network (ELKH), Konkoly Thege Mikl\'os \'ut 15–17, H-1121 Budapest, Hungary
         \and
             ELTE E\"otv\"os Lor\'and University, Doctoral School of Physics, P\'azm\'any P\'eter s\'et\'any 1/A, H-1117 Budapest, Hungary
         \and
             CSFK, MTA Centre of Excellence, Budapest, Konkoly Thege Mikl\'os \'ut 15-17., H-1121, Hungary
         \and
             Baja Astronomical Observatory of University of Szeged, H-6500 Baja, Szegedi \'ut, Kt.~766, Hungary
         \and
             ELKH-SZTE Stellar Astrophysics Research Group, H-6500 Baja, Szegedi \'ut, Kt.~766, Hungary.
         \and
             ELTE E\"otv\"os Lor\'and University, Gothard Astrophysical Observatory, Szent Imre h. u. 112, H-9700 Szombathely, Hungary
             }

   \date{Received ...; accepted ...}

  \abstract
   {}
   {Asymmetric features in exoplanet transit light curves are often interpreted as a gravity darkening effect especially if there is spectroscopic evidence of a spin-orbit misalignment. Since other processes can also lead to light curve asymmetries this may lead to inaccurate gravity darkening parameters. Here we investigate the case of non-radial pulsations as possible sources of asymmetry and likely source of misinterpreted parameters through simulations.}
   {We obtained a series of simulated transit light curves of a hypothetical exoplanet-star system constructed to study the phenomenon: a host star with no gravity darkening exhibiting small amplitude pulsations, and a typical hot Jupiter in a circular, edge-on orbit. A number of scenarios of single- or multiperiodic, radial- or nonradial pulsations of various amplitudes were considered, and a proper account of the obscuring effect of transits on all the surface intensity components was made. The magnitude of amplitude and phase modulations of nonradial pulsations during transits was also also investigated. We then fitted both a
   non-gravity-darkened, and a gravity-darkened, free spin-orbit axis model on the data. The Akaike and Bayesian Information Criteria were used for an objective selection of the most plausible model. We then explored the dependence of the parameter deviations on the pulsation properties, in order to identify configurations that can lead to falsely misaligned solutions.}
   %
  {The modulation of amplitudes of nonradial pulsations during transits have extremely low levels, so that the nonradial nature of pulsations can be safely ignored. Low-amplitude pulsations in general do not affect the determination of the system parameters beyond their noise nature. However, frequencies close to multiples of the orbital frequency ($n*f_{orb}$) are found to cause distortions leading to solutions with a side tilted stellar rotational axis, they are therefore preferable to clean beforehand for the sake of a correct analysis. Additionally, for cases with higher-amplitude pulsations, it is recommended to pre-process and clean the pulsations before analysis. 
  }
   {}


   \maketitle
%

\section{Introduction}

The large amount of exquisite quality photometric data furnished by the 
\textit{Kepler} and Transiting Exoplanet Survey Satellite (TESS) space missions (\citealp{KeplerBorucki2010Sci...327..977B, TESSRicker2015JATIS...1a4003R}),
following the pioneering \textit{CoRoT} satellite (\citealp{CoRoT2009A&A...506..411A}),
has led to the discovery of thousands of exoplanets by the transit method, confirmed later by ground-based observations using the radial velocity method. These combined data also allowed the determination of absolute physical parameters of the planets, once the properties of the host stars became known with sufficient accuracy. There is now a large database available with different exoplanet radii and masses, and on various orbits, which allowed detailed statistical studies regarding their distribution and architectures \citep[see][for a review]{Zhu2021ARA&A..59..291Z}.

Given that the inferences on planetary radii and masses require the  characterisation of their host stars at least in terms of mass and radius, there are several ongoing, dedicated systematic efforts to achieve this goal. 
Asteroseismology has proven particularly useful both in giant (e.g. \citealp{ChaplinMiglio2013ARA&A..51..353C}) and solar-type stars (e.g. \citealp{SilvaAguirreetal2015MNRAS.452.2127S}) by analysis of solar-type oscillations, the regular frequency pattern of which allows the derivation of asteroseismic masses and radii with high precision. 
The case of other pulsating star types is not so bright yet. 
$\delta$~Scuti stars, which could be the next logical step in extending asteroseismic studies, are still challenging, despite the tremendous progress in available data, also thanks to the space telescope missions mentioned above \citep{Guzik2021FrASS...8...55G}. 
Although recent progress has been achieved in the case of early type $\delta$~Scutis \citep{Beddingetal2020Natur.581..147B} and also in some of their regular counterparts \citep{Murphyetal2021MNRAS.502.1633M}, the majority still defy a definitive asteroseismic solution. They have irregular, complex frequency pattern, caused by to large, uneven rotation splittings due to their fast rotation, and the still unexplained excitation "rules" of their modes.
Additionally, the pulsation amplitudes have a large range, from 370 ppm to approximately 30000 ppm in relative flux \citep{Uytterhoeven_2011A&A...534A.125U}.
Stars of spectral type A/F, to which $\delta$~Scutis also belong, are generally excluded from radial velocity exoplanet surveys (\citealp{Calif2010ApJ...721.1467H, HARPS2020A&A...633A..44G, CalifLegacy2021ApJS..255....8R}), because their heavily rotationally broadened spectral lines make it challenging to obtain precise RV measurements.
However, although not particularly favored in dedicated transit surveys too, they may still show up in subsequent transit searches among space mission data.
Exoplanets found around such hot stars have the potential of verifying the theory of photoevaporation of primary planetary atmospheres under the harsh light of their hot young stellar hosts (\citealp{OwenWu2013ApJ...775..105O, OwenWu2017ApJ...847...29O}), conceived to explain the "radius gap" in the distribution of Kepler planetary radii \citep{RadGap2017AJ....154..109F}.
Therefore exoplanet searches directly targeting $\delta$~Scuti stars (\citealp[e.g.][]{Heyetal2021AJ....162..204H}) can be extremely fruitful.
There are also transiting systems showing asymmetric light curves. 
The first of this kind was Kepler-13 \citep[][]{KOI13_SzaboI2011ApJ...736L...4S, KOI13_SzaboII2012MNRAS.421L.122S}, explained as a combined effect of gravity-darkened stellar surface due to fast rotation and a tilted planetary orbit.
In recent years exoplanet-hosting stars exhibiting $\delta$~Sct pulsations and gravity darkening \citep[WASP-33,][]{2010MNRAS.407..507C, 2011A&A...526L..10H, vonEssen2014A&A...561A..48V, 2022ApJ...925..185D,2022A&A...660L...2K} were also investigated.
At the same time, intrinsic stellar activity, like spots, flares, pulsation or granulation may also yield asymmetric transits. Studies inspecting modelled high-precision photometric light curves of exoplanet systems with an active host star showed that the determined transit parameters -- like transit time, transit depth and transit width -- are influenced by the presence of stellar spots \citep[e.g.][]{Oshagh_2013A&A...556A..19O}. Moreover,
\citet{Oshaugh2016A&A...593A..25O} 
investigated the effect of the polluting signal of stellar activity on the spectroscopic determination of such quantities using the Rossiter-McLaughlin effect, and showed that in some cases it is capable of distorting the fit, falsely rendering an aligned system as misaligned.
%
%
Pulsations could also constitute a problem, especially when their amplitudes are too low to be properly detected and separated from light curve. It is true that when multiple individual transits are combined into an average transit curve, the same process also tends to suppress any periodic signal that is not in resonance with the planetary motion. However, when there are either frequencies close to an orbital resonance, that may have an asymmetric footprint on the transit curve when folded with the orbital period.
Another concern could be the modulation phenomenon of the non-radial pulsations subjected to transit. This phenomenon can be very pronounced in ordinary eclipsing stars \citep[e.g.][]{Kim2002A&A...391..213K, Rodriguez_2004MNRAS.347.1317R, Rodriguez2010MNRAS.408.2149R, Maceroni_2014A&A...563A..59M}. In case of transits the modulation is expected to be much less due to the much smaller size of the eclipsing body, but still could cause distortions, if the effect of multiple modes adds up to the level capable of affecting parameters that depend very sensibly on the minute details of the transit curve. Parameters of this kind are the spin-orbit angles, for example. 

To mitigate the effects of irregular stellar activity, these nuisance signals can be handled as a time-correlated noise, in a wavelet-based approach \citep{2009ApJ...704...51C}, like in the Transit and Light Curve Modeller (\texttt{TLCM}, \citealp{2020MNRAS.496.4442C, 2021arXiv210811822C}).
On the other hand, this approach is not automatically guaranteed to work in the case of multiperiodic pulsation signals, because the red component is not assumed to give rise to such asymmetries. 

The goal of our paper is the exploration of the effect of small-amplitude, multiperiodic, pulsations on the determination of planetary system parameters from transit light curves.
We consider two scenarios for each of them. The first scenario features single period, radial-only pulsations. In the second scenario we examine the effect of the modulation of the multiperiodic, non-radial pulsations on the parameters, handled either as radial mode harmonics subjected to eclipses or as red noise. For each case we run a full light curve analysis, and aim to identify configurations which are particularly vulnerable to such polluting signals. 
In Sect.~\ref{sec:method} we describe the methods used for this investigation.
Sect.~\ref{sec:result} presents the results and their interpretations.
In Sect.~\ref{sec:disc} we discuss them in a broader context, regarding implications on future transit light curve analysis procedures.
In Sect.~\ref{sec:conc} we summarise our findings.


\section{Methods}\label{sec:method}

We considered a hypothetical exoplanet-star system with a pulsating host star without rotational distortion, thus having uniform equilibrium surface brightness with no gravity darkening and no particular symmetry axis; and with various pulsation scenarios, both with single period radial and multiperiodic nonradial pulsations. Artificial light curves were synthesized using two separate tools: one for computing the transits of the time-independent, uniform stellar surface, and a second one for doing the same for the time-dependent pulsating surface brightness component. The two variations were added together and supplemented with an artificial noise. Then we used a third modelling tool to fit both gravitationally undarkened and darkened transit models on the data. The latter models also included the spin-orbit angles as fitting parameters. We then identified the cases leading to misaligned configurations. All the steps are presented below in detail.

\subsection{Synthetic light curves}
We studied the effects of a transit on the pulsational pattern in a generic model that was based on the pixel-level in-silico visualisation of the pulsating stellar surface while it was occulted by the planet. The local intensities were summed up pixel-by-pixel, leading to a synthetic brightness of the star, when it exhibited a certain pulsational pattern that was partially occulted by the planet at a certain position along the transit chord. This technique enabled an extensive analysis of the different pulsation modes, and we could identify which of these are the most sensitive to planetary transits, i.e. which are the modes where planetary transits most commonly cause virtual amplitude and phase modulations due to a transiting companion.

 We wanted to avoid fallacies caused by so-called inverse crimes, when the same model used both for simulating the data and then again for fitting them results in a much better recovery of the input parameters than in real applications, leading to overoptimistic conclusions. Therefore we used one set of tools for creating the synthetic light curves \citep[\texttt{fitsh/lfit}; \texttt{pulzem}, see][]{fitsh2012MNRAS.421.1825P,Biro2011MNRAS.416.1601B}, and another tool for the modelling \citep[\texttt{TLCM}, see][]{2020MNRAS.496.4442C}.

 The light variation due to the transits of the static surface brightness component was modelled using the \texttt{fitsh} package, which uses analytical formulae \citep{2002ApJ...580L.171M} to compute a resolution-independent light curve.
 Its lack of capability to model rotationally distorted, gravity-darkened stars did not pose any disadvantage because our input configuration was just a slowly rotating, undistorted star with no gravity darkening.
One of the input parameters of \texttt{fitsh} is a combination of the period $P$, relative semi-major axis $a/R_\text{s}$ and conjunction parameter $b$:
\begin{equation}
    \omega = \frac{2\pi}{P}\frac{a}{R_s}\frac{1}{\sqrt{1-b^2}}.
\end{equation}

The photometric dataset was modelled with a quadratic limb-darkening law with coefficients $\mu_1=0.209$ and $\mu_2=0.216$ chosen during preliminary tests and kept fixed throughout the investigation.

The light variation due to the pulsations was generated with the forward modelling utilities of the \texttt{pulzem} package \citep{Biro2011MNRAS.416.1601B}. Created to perform eclipse mapping of pulsation patterns in eclipsing binaries, it properly takes into account the partial occultation of the surface pulsation patterns, and handles planet-sized eclipsing objects too. 
The geodesic grid scheme based on \citet{Hendry1992ApJ...388..603H} was implemented, involving a sub-triangulation of the faces of an icosahedron inscribed into the stellar volume and then radially projecting the resulting grid onto the surface. The procedure yields a nearly uniform surface resolution with the least number of triangular tiles.
In addition, the intersection of the eclipser's shadow - a circular disc in our case - with the triangular tiles is exactly taken into account.
The usual features of limb darkening, rotational distortion and gravity darkening of the host star are also properly implemented -- although in our case the latter two were not required in the forward modelling.
We gradually increased the grid resolution until the change in the normalised flux of the transit curve decreased below the level of the uncorrelated error component.
 The condition was fulfilled at the value of $15$ for the sub-triangulation parameter, i.e. each side of the icosahedron is divided into 15 segments,
giving $15^2=225$ tiles for each of the $20$ faces, and resulting in $20\cdot 15^2=4500$ tiles for the whole stellar surface, half of which is actually visible for a stellar disc. Rotation of the static pattern is not required, while it only requires a slightly modified frequency for the non-linear pulsation patterns, so the stellar surface actually needs no rotation at all.
The same limb darkening used for the static light curve was applied to the pulsation surface patterns too, before generating the combined pulsation flux of all modes.

The reason behind using one tool for synthesizing the static flux and another one for the pulsating flux component is that there is no suitable tool available yet capable of computing both components with the same accuracy. \texttt{fitsh} is virtually exact for transit light curves, but cannot submit arbitrary time-dependent surface patterns to the same transit computation. pulzem computes both, but its accuracy is intrinsically limited by the rasterization approach. Static transit light curves of \texttt{fitsh} and \texttt{pulzem} showed differences that could not be diminished by increasing the surface resolution beyond all limits. Although the pulsation patterns suffer from the same relative inaccuracies, their small amplitudes explored in this study, of a few percents of the transit depth, render these errors below the applied artificial uncorrelated noise component. Therefore the use of this combined simulation method is appropriate.

\begin{table}
\caption{The geometric and physical properties of the test system, shown as the two equivalent parameter sets
passed to the two modelling tools.}            
\label{table:2_parsv2}      
\centering                          
\begin{tabular}{c c c}        
\hline\hline                 
& Parameter      &  Value \\    
\hline                        
                & $P$                       & 1.22 d \\
                & $b$                       & 0.     \\
\texttt{fitsh}  & $\omega$                  & 18.975 \\
                & $R_\text{p}/R_\text{s}$   & 0.1117             \\
                & $\mu_1$                   & 0.209\\
                & $\mu_2$                   & 0.217\\
\hline
                & $R_\text{s}/a$            & 0.2714 \\
                & $R_\text{p}/a$            & 0.0303 \\
                & $i$                       & 90.\\ 
\texttt{pulzem} & $e$                       & 0. \\
                & $N$                       & $20\times15\times15$ \\
                & $\mu_1$                   & 0.209\\
                & $\mu_2$                   & 0.217\\
\hline                                   
\end{tabular}
\end{table}

The first investigated scenario was a test system consisting of a spherical star and a planet with a relative size $R_\text{p} \sim$ 0.1 $R_\text{s}$, orbiting its host star with a period of about 2~days. This is a typical hot Jupiter configuration. The orbit was completely circular and seen edge-on (inclination is 90 degrees).
Uncorrelated random noise was added to the light curve. Its level was chosen to match the error of an average target of ~10-11~mag observed with the TESS mission, $\sim 0.0002$ in the normalised flux.
The modelled transits had a time duration of 0.1 orbital phases, and the depth of the transit in was approximately 0.013. Table~\ref{table:2_parsv2} summarises the input parameters of the two tools used for the simulation. Each parameter set was used for different investigation as written in the following subsections.

For the second scenario, we used an identical system with the only difference that an inclination value of $89.89^\circ$ was used, corresponding to the WASP-33 system which ultimately inspired this work.

\subsection{Modelling the pulsations}

\begin{figure}
   \centering
   \includegraphics[width=\hsize]{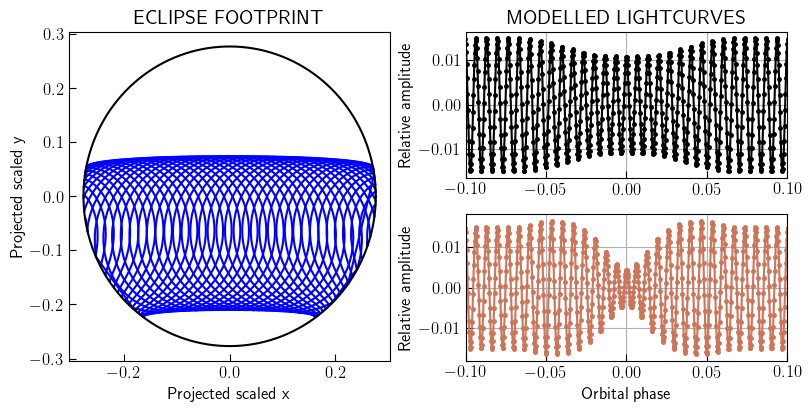}
      \caption{Modulations of pulsations by the eclipse computed in a hypothetical binary system. Both stars are spherical, the orbit is circular, the inclination is 85$^{\circ}$, and the secondary radius is half of the primary radius. The left panel shows with blue lines the eclipse footprint of the secondary on the circular disc of the primary. Only the footprints corresponding to every fifth point of the light curves on the right panels are shown for clarity. The right panels show the actual light changes for two selected modes: (0,0) on top, and (2,1) on bottom.}
         \label{fig:0a-intro}
   \end{figure}
   
   The surface intensity pattern of the various small amplitude non-radial pulsations in stars with no significant rotational and tidal distortion can be described by spherical harmonics $Y_\ell^m(\theta, \varphi)$, with degree $\ell$ and azimuthal order $m$ \citep{2010aste.book.....A}. The integrated flux coming from an uneclipsed, fully visible stellar disc naturally does not contain any information on the nonradial modes of the pulsations. During eclipses (or transits, in exoplanetary parlance) an area moving through the stellar disc is excluded from the integration, leading to a modulation of the periodic signals. The instant amplitudes and initial phases will show variation patterns that depend on the geometric configuration and also on the particular non-radial mode, labelled by the horizontal mode numbers $\ell$ and $m$ in the context of the spherical harmonics approximation. For eclipsing binary systems with similar stellar radii this can be significant (see Fig.~\ref{fig:0a-intro} for examples), in extreme cases leading to the appearance of "hidden modes" during the eclipses. %
   As a consequence, in the absence of mode identification, without information on the  modulation patterns, the proper disentangling of pulsations and eclipse flux variation becomes a challenge. Iterative procedures are involved to tackle the problem \citep[e.g.][]{Prsa2005ApJ...628..426P,Conroy2020ApJS..250...34C}. 
   They are lengthy and cumbersome, but rewarding in the correct determination of the absolute parameters.
   For planetary transits with much smaller eclipsed regions, the modulations themselves are not expected to be of concern. Nevertheless we have investigated their magnitudes.

For the first of the scenarios mentioned above we assumed the presence of single period, radial pulsations. For the second scenario we assumed various multiperiodic, multimode, non-radial pulsations, resembling the ones expected in $\delta$~Scuti type variables.
The corresponding synthetic time series were computed in all cases using \texttt{pulzem}, as described in the previous subsection.

Table~\ref{tab:2_freq_simp} lists the properties of the pulsation signals used in the single radial mode case.
These refer to the contribution of each pulsation mode to the integrated flux \textit{outside the transits}, according to the formula
\begin{eqnarray}
   f(t) = A \cdot \sin\left(2\pi\left(f t + \varphi\right)\right).
   \label{eq:sinform}
\end{eqnarray}
The forward modelling tool of \texttt{pulzem} scales the surface patterns and shifts their initial phases to yield the given parameters, before subjecting them to the eclipses.

\subsubsection{Single mode, radial pulsations}

We used the case of single, radial pulsations to test whether handling them as red noise is appropriate, i.e whether the fitting procedure still yields parameters close to their input values. 
Red noise was introduced in \texttt{TLCM} primarily to cope with the signatures of stellar activity, for which it proved to be excellent \citep{2022arXiv220801716K}.

\subsubsection{Multiperiodic, nonradial pulsations}

We also tested how much bias can be expected from non-radial pulsations in various geometrical configurations. It is evident that the leading parameter is the relative exoplanet radius: larger radii influence larger areas on the stellar surface and thus cause larger modulations.
Therefore an upper limit can be determined for the possible detection.

Generally we wanted to map the variation of the light-curve modulations with the alignment angle of the exoplanet orbit.
Therefore we created a sequence of systems by varying the projected spin-axis angle $\lambda$ between 0 and 180 degrees, in steps of 15 degrees. This also changes the projection of the surface pulsation patterns on the apparent stellar disc with respect to the transit trajectory and also the type of modulation as a consequence. We assume that the symmetry axis of the pulsations coincides with the axis of rotation, as usual.

\texttt{pulzem} has an option to compute the instantaneous amplitude and initial phase of each mode in every moment, and output them directly as amplitude and phase modulation data.
These quantities are mode number specific, but do not depend on other characteristics of the pulsation, therefore we used one fiducial pulsation with fixed amplitude, frequence and initial phase for all modes.
We modelled different ($\ell$,$m$) modes in each system, including the radial mode (0,0) which serves as the base modulation relative to which all the others are evaluated.
Even the amplitude of a radial mode suffers a slight decrease due to the transit obscuring part of the disc, but no phase modulation happens, as the whole surface oscillates coherently.

\begin{table}
	\centering
	\caption{Data of pulsations used for single-mode, radial mode cases. The frequencies are in units of the orbital frequency. The initial phases $\varphi$ were chosen randomly in the interval $[0,1]$. 
	The amplitude was uniformly $0.0006$. The IDs of the models were composed from the frequency and initial phase numbers, with the simple model id initial "i". E.g. "i02" used the first frequency and the third initial phase values (counting starts from zero).}
	\label{tab:2_freq_simp}
	\begin{tabular}{c||c} %
		\hline
		$f [f_\text{orb}]$ & $\varphi$ \\
		\hline
             0.25652 & 0.78636 \\
             0.31001 & 0.31961 \\
             0.54229 & 0.43413 \\
             3.24241 & \\
             3.21314 & \\
             2.58046 & \\
            13.21510 & \\
            26.10723 & \\
            19.87613 & \\
            19.37048 & \\		
		\hline
	\end{tabular}
\end{table}

As we progressed with our study, we aimed to analyse a more realistic model containing multiple frequencies. 
The amplitudes attributed to the modes are small compared to typical $\delta$~Sct pulsations, because they were inspired by the WASP-33 system, and for this work we assumed some incidental ($\ell$,m) nonradial mode numbers (there is no known mode identification presented in literature for this system). 
The summary of frequencies and their inclusions in the \texttt{nXX} model series in this paper can be found in Table~\ref{table:1:nfreq}. In our final set of simulations and analyses, we utilized the same frequencies as in the case of \texttt{n02} but increased their amplitudes by a factor of 10. These simulated scenarios are labelled as \texttt{pXX}.

\begin{table*}
\caption{The frequency sets used in the \texttt{nXX} model series. The frequencies are given in units of the orbital frequency as usual,
the amplitude $A$ and initial phase $\phi$ are those used in the sine formula of Eq.~(\ref{eq:sinform}). Bullets ($\bullet$) in the \texttt{nXX} columns mark the frequencies included in that particular model.
The last column shows the mode numbers ($\ell$,m) assigned to each frequency.
}
\label{table:1:nfreq} 
\centering 
\small
\begin{tabular}{c|ccc|cccccc|c}
\hline
\hline
 & f & A  & $\varphi$  & \texttt{n00} & \texttt{n01} & n\texttt{02} & \texttt{n03} & \texttt{n04} & \texttt{n05} & ($\ell$,m)\\
f\_id & ($f_\text{orb}$) & ppm &  & &  & & & &  &  \\
\hline
F1 & 30.0047  & 670  &  0.7695  &   & $\bullet$ & $\bullet$ & $\bullet$ & & & (1,-1) \\ 
F2 &  2.8300  & 667  &  0.5968  & $\bullet$ & $\bullet$ & $\bullet$ &   & & $\bullet$ & (0,0)\\ 
F3 & 14.6532  & 557  &  0.5714  &   &   & $\bullet$ & $\bullet$ & $\bullet$ & $\bullet$ & (1,1)\\
F4 & 31.3453  & 496  &  0.4724  &   & $\bullet$ & $\bullet$ & $\bullet$ & $\bullet$ & $\bullet$ & (1,1)\\
F5 &  3.6987  & 437  &  0.5490  & $\bullet$ & $\bullet$ & $\bullet$ &   & & $\bullet$ & (1,1)\\
F6 &  0.3586  & 337  &  0.8244  & $\bullet$ & $\bullet$ & $\bullet$ &   & & $\bullet$ & (2,2)\\
F7 & 11.2085  & 327  &  0.1155  &   &   & $\bullet$ & $\bullet$ & $\bullet$ & $\bullet$ & (1,1)\\
F8 & 30.5633  & 290  &  0.8003  &   & $\bullet$ & $\bullet$ & $\bullet$& $\bullet$ &   $\bullet$ & (2,1)\\
F9&  1.4248  & 243  &  0.3316  &   &   & $\bullet$ & $\bullet$ & $\bullet$ & $\bullet$ & (2,1)\\
F10&  0.1519  & 247  &  0.7518  & $\bullet$ &   & $\bullet$ &  & &  $\bullet$ & (2,1)\\
\hline
\hline
\end{tabular}
\end{table*}

\subsection{Performing fits with TLCM}

We made use of the Transit and Light Curve Modeller code \citep[\texttt{TLCM};][]{2020MNRAS.496.4442C} to solve the combined synthetic light curves of the simulated stellar pulsations and the transits. The transit modelling 
in \texttt{TLCM} is done via the \cite{2002ApJ...580L.171M} model, which is parameterized by the orbital period, $P$, the time of mid-transit, $t_\text{C}$, the star-to-planet radius ratio, $R_\text{p}/R_\text{S}$, the scaled semi-major axis, $a/R_\text{S}$, and the impact parameter, $b$. Stellar limb-darkening was taken into account with a quadratic limb-darkening formula, described by $\mu_1$ and $\mu_2$. \texttt{TLCM} also incorporates the wavelet-based routines of \cite{2009ApJ...704...51C} for handling the time-correlated noise which can be made up of both instrumental and astrophysical effects (including flares, granulation, stellar pulsations, etc.). The noise is characterized by $\sigma_r$ (for the red component) and $\sigma_w$ (for the white component). The wavelet-based method for handling red noise of \texttt{TLCM} was tested in \cite{2021arXiv210811822C, 2022arXiv220801716K} and was found to be consistent. In our case, the correlated noise is made up solely of stellar pulsations. \cite{2022A&A...660L...2K} reported that the wavelet-based approach of \texttt{TLCM} can handle the $\delta$~Scuti type pulsations of WASP-33.

The uneven surface brightness of the host star due to the gravity darkening effect induced by its rapid rotation may yield asymmetric light curves in certain configurations \citep{2009ApJ...705..683B,Barnes2011ApJS..197...10B,Ahlers2019AJ....158...88A}. We used a slightly modified version of the gravity darkening model of \texttt{TLCM} \citep{2020A&A...643A..94L, 2021arXiv210811822C}, incorporating a full Roche-geometry \citep{1979ApJ...234.1054W} which is used for the calculation of the exact shape of the stellar surface. The modelling takes the projected stellar rotational velocity ($v \sin I_\star$) into account when calculating the stellar surface potential. This is then used to describe the surface temperature variations in combination with the gravity darkening exponent $\beta$. There are two additional fitting parameters: the inclination of the stellar rotational axis $I_\star$ (measured from the line of sight) and the projected spin-orbit angle $\lambda$.

For every synthesized light curve we performed at least two series of fits.
The first series corresponds to the input model configuration: no gravity darkening, aligned star configuration ($I_\star=90^\circ$, $\Omega_\star=90^\circ$ or $\lambda=0^\circ$).
A second series assumed nonzero gravity darkening using a temperature gravity darkening coefficient $\beta=0.25$ and a fixed $v \sin I_\star =$ 86.5 km/s, and both stellar angles being fitting parameters. 
We note that with this choice the amount of gravity darkening from the polar to the equatorial zone is completely determined and thus its effect on the transit light curve is fully characterised by the two angles and the conjunction parameter $b$.
Sporadically we also fitted the conjunction parameter in order to assess its influence on the solution in particular cases.
The scaled semi-major axes, $a/R_\text{S}$, star-to-planet radius $R_\text{p}/R_\text{S}$, mid-transit time $t_C$, period $P$ were always fitted in our analysis as well as the red noise $\sigma_r$ and white noise $\sigma_w$. The limb-darkening parameters $u_+=\mu_1 + \mu_2$ and $u_-=\mu_1 - \mu_2$, as required by \texttt{TLCM}, were kept fixed.

For an objective selection of the best model we made use of the Akaike and the Bayesian Information Criteria \citep[$AIC$ and $BIC$, respectively;][]{https://doi.org/10.1002/wics.1460}, defined as
\begin{align}
    AIC &= 2 \cdot n_\text{par} + n_\text{obs} \cdot \log \left(RSS /n_{obs}\right),  \\
    BIC &= n_\text{par} \cdot \log n_\text{obs} + 
            n_\text{obs} \cdot \log \left(RSS / n_\text{obs}\right),
\end{align}
where $n_\text{par}$ is the number of parameters, $n_\text{obs}$ is the number of observations, and $RSS$ is the sum of squared residuals. The only difference between the two formulae is that the penalty factor for the number of model parameters increases steeper for the BIC due to the $\log n_\text{obs}$ factor.
The model with the lowest $AIC$/$BIC$ values is the most plausible, while the relevance of the other models is measured in terms of the difference of their given $AIC$/$BIC$ from the lowest one. If this difference $\Delta AIC$ or $\Delta BIC$ is smaller than 3, then the empirical support for the model is substantial. For differences between 4 and 7 the plausibility is considerably smaller, and above 10 the model is implausible.

Due to the pulsations being fitted in the form of red noise, we computed two sets for both criteria. In one of them the red noise is accounted for in the model, which is labelled 'yn' (i.e. 'yes noise'). In the other one, the red noise is excluded from the model, instead it is considered just another addition to the residuals, hence it is labelled 'nn' (i.e. 'no noise'). 

In the case of \texttt{pXX} series the enhanced pulsation amplitude might imply a red noise component so large that it approaches the limit of analytic capability of \texttt{TLCM}. Whenever this happened, whe identified the frequencies using \texttt{Period04} \citep{period042005CoAst.146...53L}, and subtracted them from the datased as pure harmonic signals. After having ensured that the prewhitening process is satisfactory, the \texttt{TLCM} analysis was repeated. 

\section{Results}\label{sec:result}

\subsection{Single mode, radial pulsation}\label{sec:singlemodes}

\begin{figure*}
   \resizebox{\hsize}{!}
            {\includegraphics[width=0.7\textwidth]{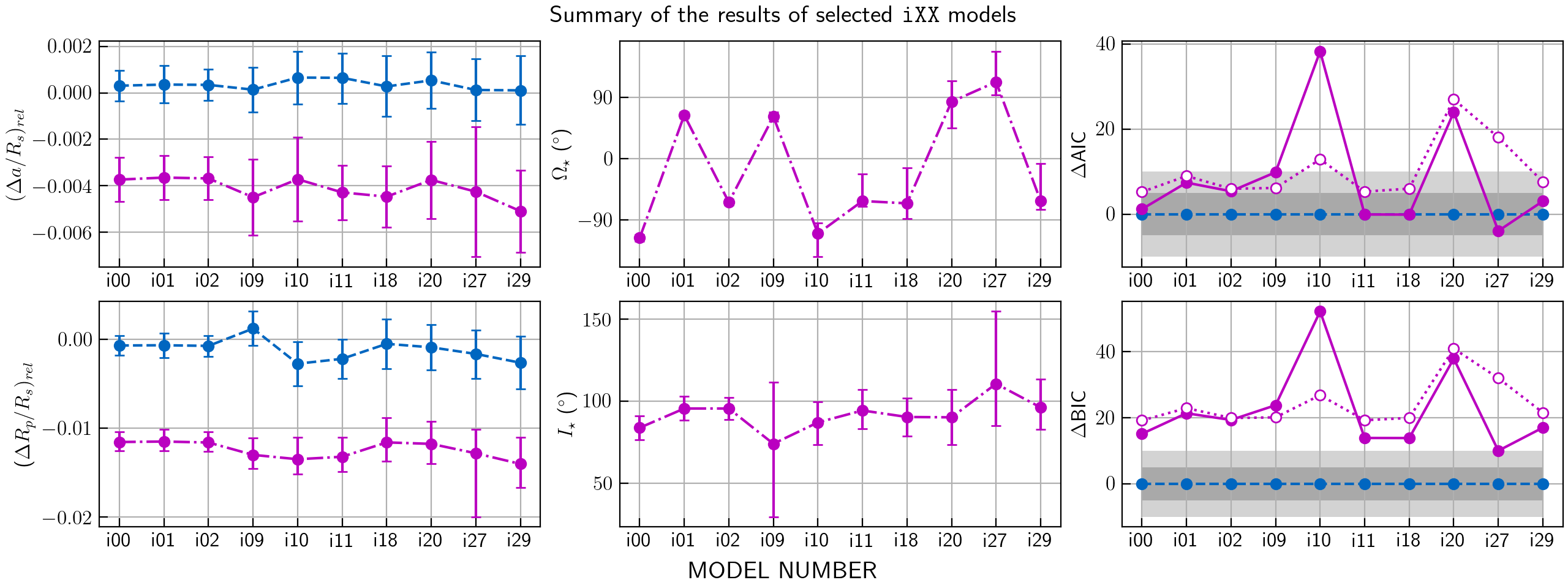}}
      \caption{Summary of the results for selected \texttt{iXX} models. Blue colours mark the aligned case with $\beta=0$ and magenta colours mark the free angles cases with $\beta=0.25$ in all graphs. \textbf{Left panel:} the relative deviations of $a/R_S$ and $R_p/R_S$ fits from their input values. \textbf{Middle panel}: the fitted angles $\Omega_\star$ and $I_\star$ free angles case assuming $\beta = 0.25$ \textbf{Right panel:} the computed differences $\Delta AIC$ and $\Delta BIC$ between the free angles and fixed angles cases. Filled circles mark the 'yn' scenarios (red noise considered), empty circles mark the 'nn' scenarios (no red noise). The two grey shaded zones correspond to differences of 5 and 10 respectively, representing two thresholds of rejection.
              }
         \label{fig:03_res_i}
   \end{figure*}

\begin{figure*}
\resizebox{\hsize}{!}
        {\includegraphics[width=0.7\textwidth]{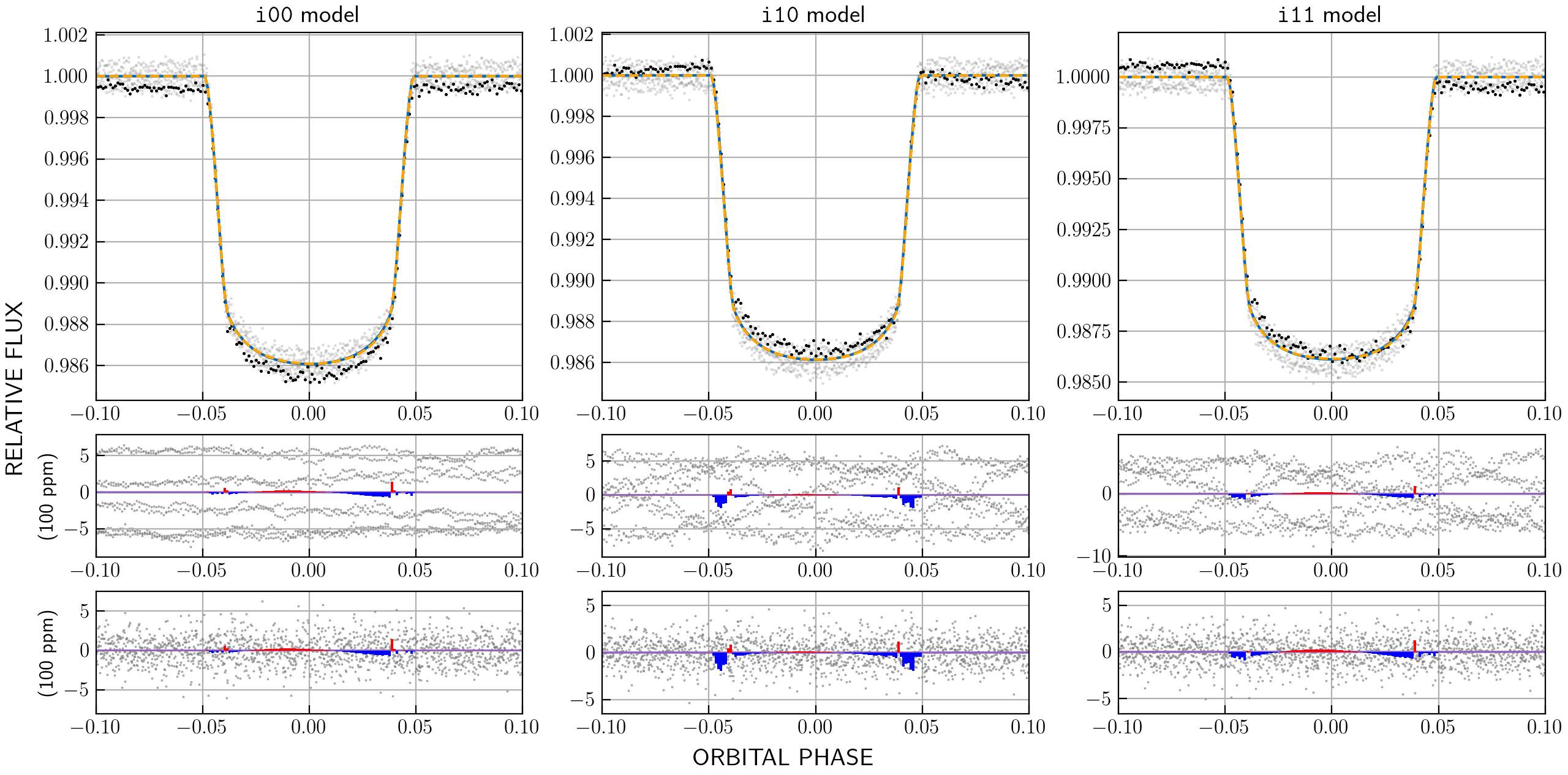}}
  \caption{Sample fits and residuals of three simulated systems from the \texttt{iXX} series. 
           \textbf{Top panel}:
           Folded light curves with grey points, with simulated transit curve samples for the first orbital period, including noise and pulsation with black circle. There are also the fits of the two models: the fixed angle, $\beta=0$ model with blue dashed line, and the free angles, $\beta=0.25$ model with orange solid line. The two fits are virtually indistinguishable on these graphs. 
           \textbf{Middle panel}: the residuals in the red noise case.
           \textbf{Bottom panel}: the residuals in the white noise case.
           The colours along the horizontal line at y = 0 
           encode the sign of the difference between the predicted fluxes of the fixed angles ($\lambda = 0$, $I_\star = 90$; at $\beta=0$) and free angles ($\beta = 0.25$) models. Positive difference is shown in red, negative in blue; otherwise it is magenta.
          }
     \label{fig:01:iXX}
\end{figure*}

The presence of single frequency pulsations did not pose any difficulty for the wavelet algorithm of \texttt{TLCM} responsible for handling the noise. The original input parameters were successfully recovered within errors in all the 30 systems, as it is shown in Fig.~\ref{fig:03_res_i} for a selection of cases. The increased scatter in the data translates into larger uncertainties in all the fitted parameters, as usual (left and middle panels of Fig.~\ref{fig:03_res_i}. 
The free axis models also yield essentially the aligned solution in the
majority of
cases, although with generally
larger $AIC$ and always larger $BIC$ values, which reassures the higher plausibility of the simpler model. There are also small deviations in the obtained relative planet radii (left panel of Fig.~\ref{fig:03_res_i}), which however still fall well within their uncertainties obtained for the pulsation-free case.
Fig.~\ref{fig:01:iXX} shows the achieved fits to the data for three of them: \texttt{i00}, \texttt{i10} and \texttt{i11}, together with the residuals shown separately for the red noise and the residuals respectively. Note that the aligned and free axis models, shown in blue and orange, are virtually indistinguishable in this setup.

\subsection{Multimode, nonradial pulsations}

\begin{figure}
   \centering
   \includegraphics[width=9cm]{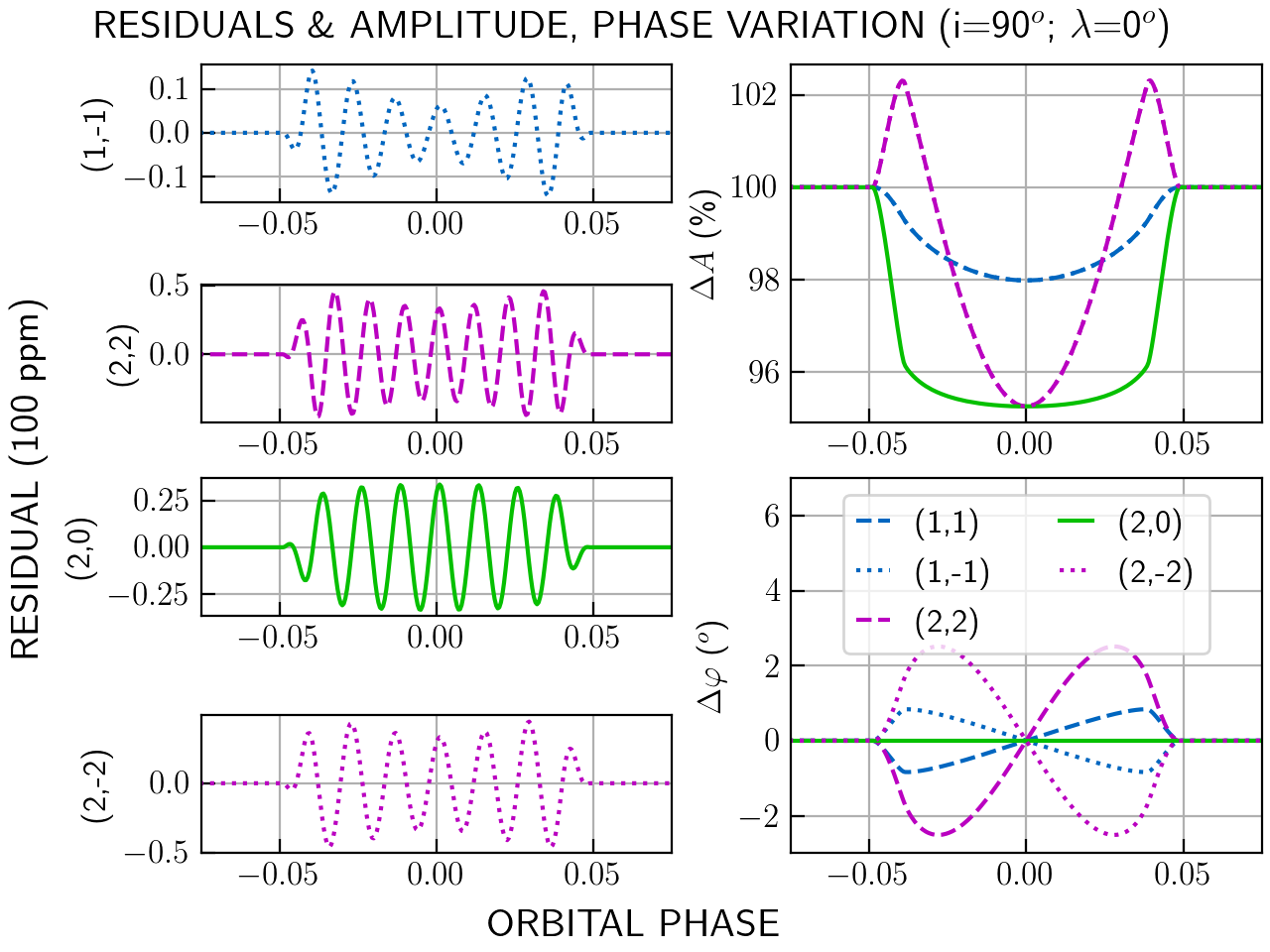}
  \caption{
           An example for modulations of some nonradial pulsations during the exoplanet transit phases in a system with aligned orbit ($i=90^\circ$, $\lambda=0^o$).
           The four panels on the left show the residuals in the signals after the subtraction of mode (0,0).
           The panels on the right show the variations in the amplitudes (top) and phases (bottom) of all visible nonradial modes. Note the small vertical scale of the phase modulation diagram, implying that such a change will be barely noticeable.
          }
     \label{fig:02:apvarintro}
\end{figure}

Fig.~\ref{fig:02:apvarintro} illustrates the modulations experienced by the integrated flux of some nonradial modes during the transits in our artificial system. The simulated light curves reveal very small apparent modulations of the "transited" pulsations for all cases. When compared to ordinary eclipsing binaries (Fig.~\ref{fig:0a-intro} in Sec.~\ref{sec:method}), the difference is especially striking, because the amplitude variations due to the transits are not larger than about 2-5 percents of the unperturbed value, and the phase variations are also of a mere 2~degrees at most. By contrast, typical amplitude modulations in eclipsing binaries stars are 20-50 percents, while the phase variations commonly reach 30-50 degrees. When this phenomenon is applied to realistic scenarios of transiting exoplanets, the modulations are much smaller than the measurement errors, therefore virtually undetectable.

For completeness we also investigated the dependence of the modulations of the same nonradial modes on the $\lambda$ projected spin-orbit angle while keeping $I_\star$ at 90$^\circ$. We obtained an almost unnoticeable amplification of the modulations.
Although we did not consider all the possible oblique configurations, the small modulations suggest that they would not be significant for any other case either.

\begin{figure*}
   \resizebox{\hsize}{!}
            {\includegraphics[width=0.7\textwidth]{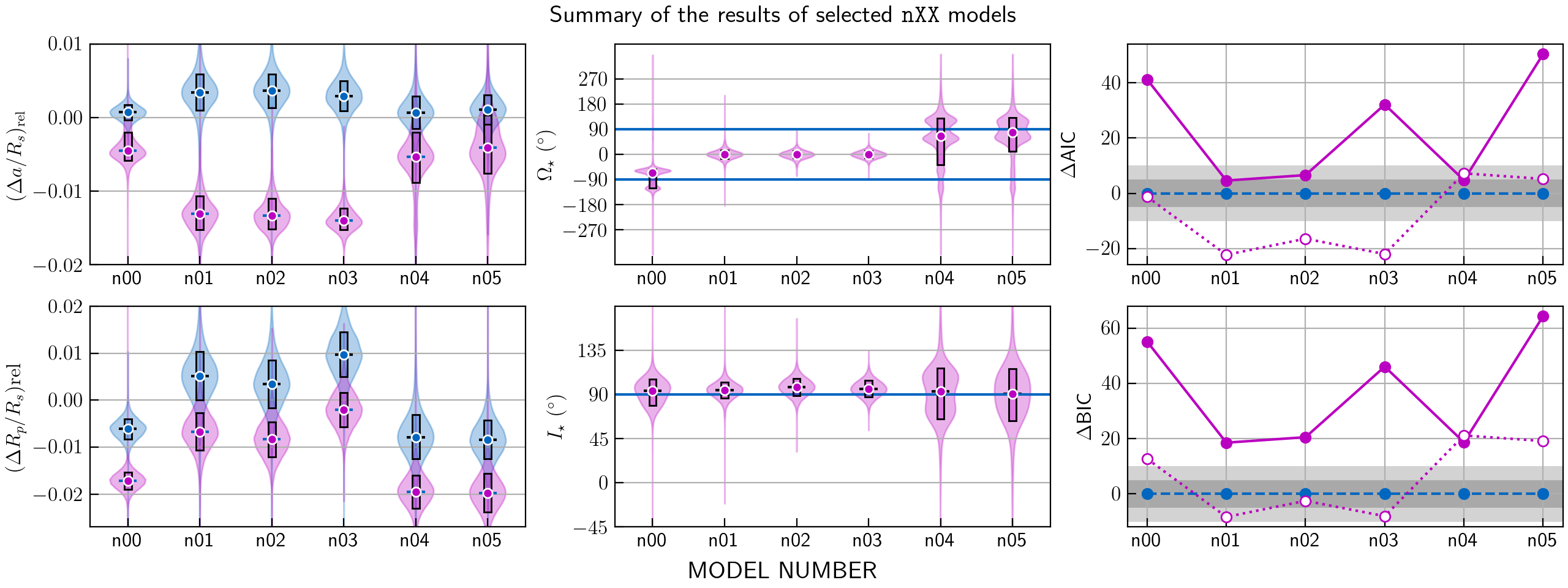}}
      \caption{Summary of the results for the \texttt{nXX} cases. The structure of the figure is similar to Fig.~\ref{fig:03_res_i}, but on the left and middle panel the marginal distribution of the corresponding parameter is depicted with so called violin plot. The circles and bars are belonging to the found median values and the uncertainty ranges.
              }
         \label{fig:03_res_n}
   \end{figure*}

Therefore it seems that the distortions in the modelled parameters due to pulsations do not depend on their radial or nonradial nature.
As a downside, neither are they useful in assessing any obliquity of the host star. As outlined in the previous section, single-mode pulsations are not a problem. Multiple pulsations on the other hand may still lead to a notable distortion of the transit curve profile if they add up under an unfortunate constellation. We have explored this possibility in the \texttt{nxx} series of trial fits.
Table~\ref{table:1:n00} and Fig.~\ref{fig:03_res_n} summarise the investigated cases and their results.

\begin{figure}
   \centering
   \includegraphics[width=9cm]{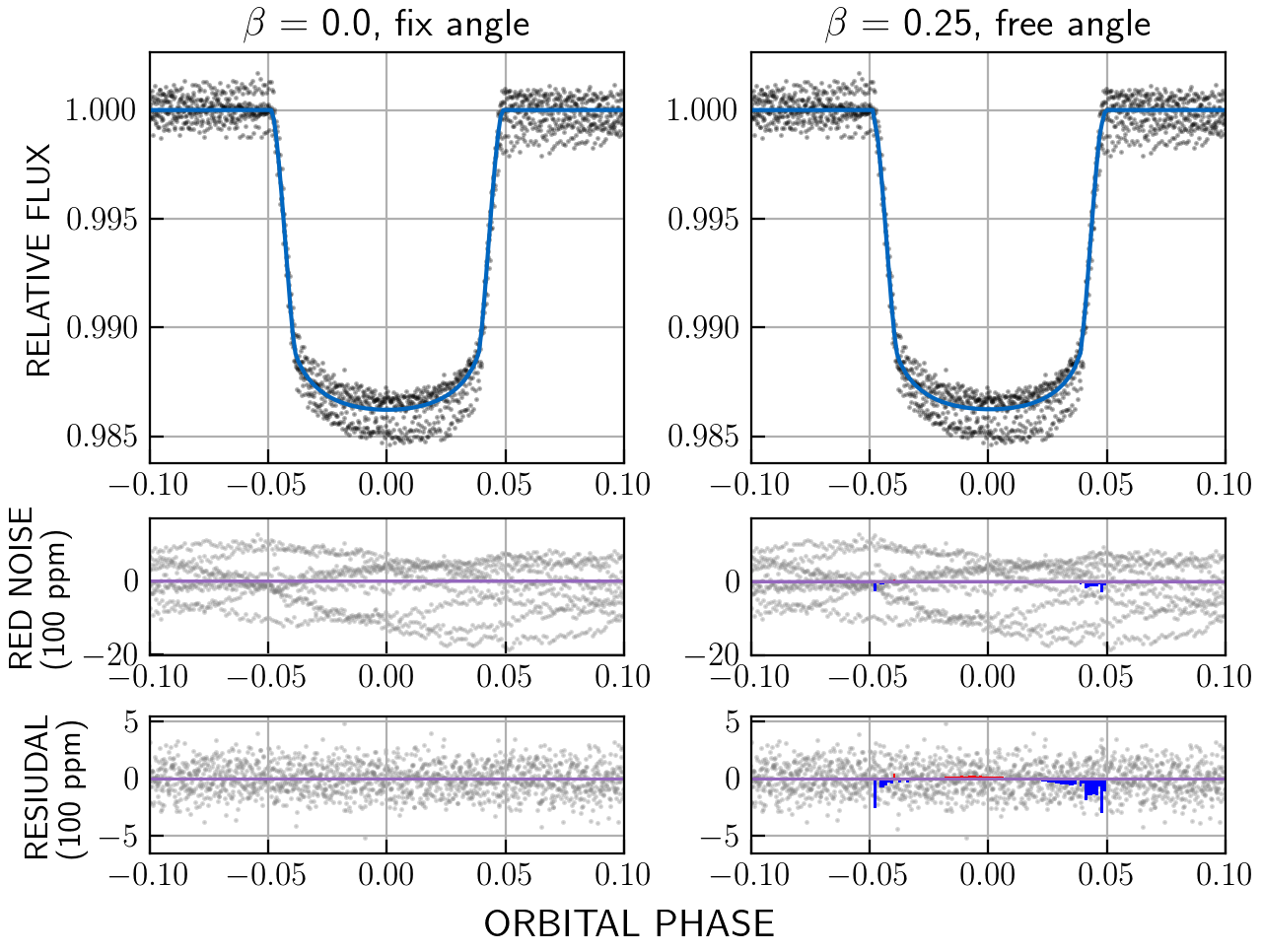}
        \caption{
            Model fits achieved for test case \texttt{n00}. The models are shown in the column titles. 
            The top row shows the synthetic data with black symbols and the modelled curves with blue lines.
            The middle and bottom rows show the fitted red noise and the final residuals, respectively.
            The coloured areas show the differences between the fits of the actual model and the leftmost model (i.e. aligned, fixed angle). Red and blue correspond to positive and negative discrepancy, respectively, and the regions with no difference are coded with magenta.
        }
         \label{fig:01:n00}
   \end{figure}

\begin{figure*}
\resizebox{\hsize}{!}
    {\includegraphics[width=0.8\textwidth]{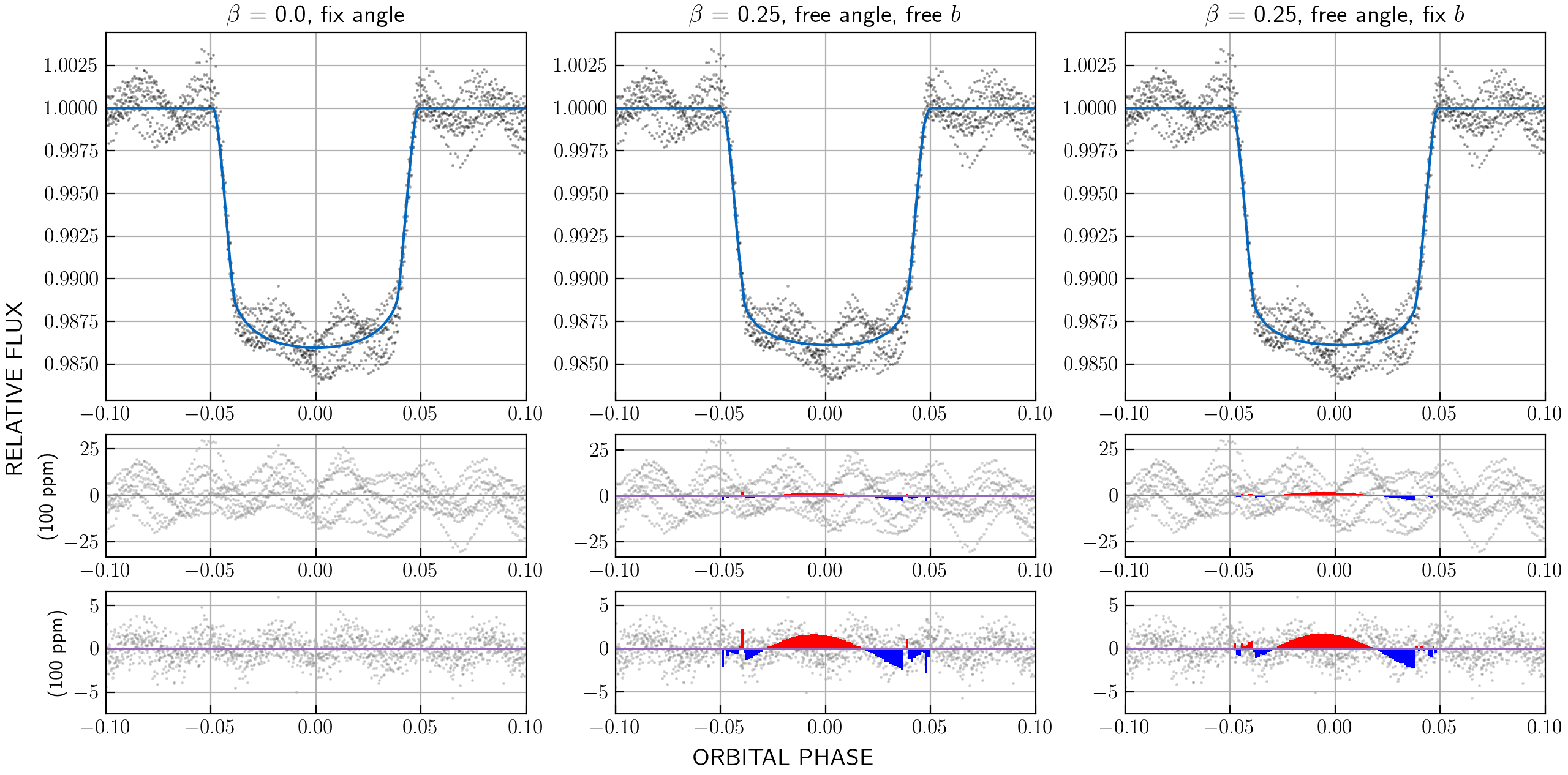}}
    \caption{%
        Model fits achieved for test case \texttt{n02}. 
        The structure and notations are the same as in Fig.~\ref{fig:01:n00}.
    }
     \label{fig:01:n02}
\end{figure*}

The first case, \texttt{n00}, yielded the correct configuration for the gravity-darkened model. Fig.~\ref{fig:01:n00} shows the achieved fits to the dataset. But a second case, \texttt{n01}, with similar randomly picked modes, gave a side tilted configuration, as did case \texttt{n02} too (see Fig.~\ref{fig:01:n02}), where all the modes have been included. 
However the associated 'yn' $\Delta BIC$ differences are in all cases large enough to signal a significantly lower plausibility of the gravity-darkened models with respect and in favour of the simpler model with fixed angles.

\begin{figure}
   \centering
   \includegraphics[width=9cm]{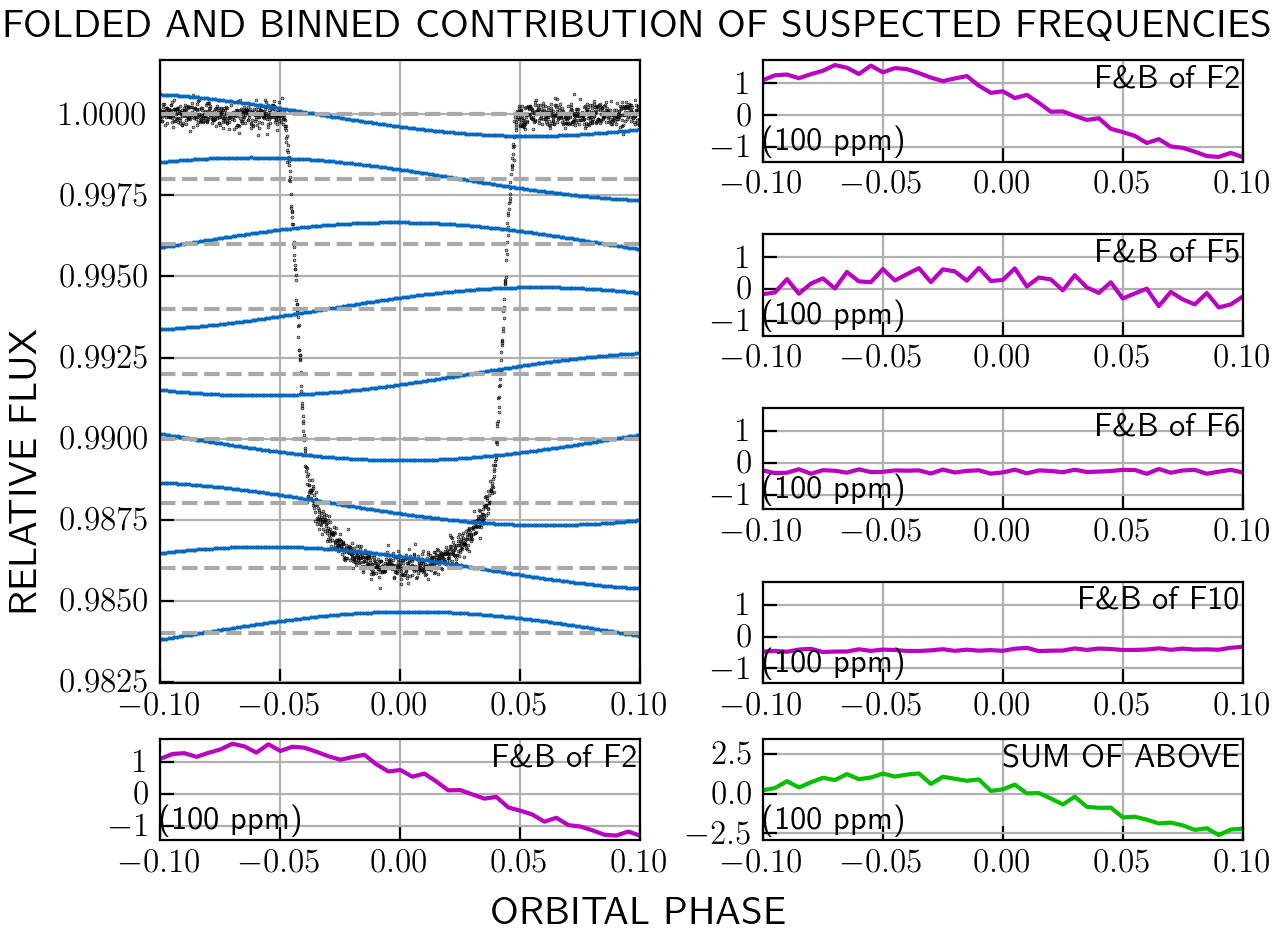}
      \caption{The contributions of some of the sinusoidal signals to the total flux around the transit phases. \textbf{Top left}: the contributions of F2 around the individual transit events with blue symbols and shifted vertically from each other for a better view,  overlaid on the folded transit curve. \textbf{Bottom left}: the contribution of F2 after folding with the orbital period and binning, shown on a scale of 100~ppm. \textbf{The right panels} show the contributions of F2, F5, F6 and F10, as well as their sum.}
         \label{fig:04_freq}
   \end{figure}

 Nevertheless, it is still instructive to explore the nature of distortions induced by the pulsation signals that governed the solution away from the aligned configuration. To do so, we made a series of trial runs in order to identify modes that contribute significantly to the related asymmetry in the transit curve. Four suspect modes were identified, corresponding to frequencies 2, 5, 6 and 10 of Table~\ref{table:1:nfreq}.
They have been selected based on the resemblance of their combined contribution to the difference in the fitted curves of the gravity-darkened and the undarkened models, that discrepancy being responsible for the tilted configuration achieved by the former model. Details of the individual as well as the combined contributions of these frequencies are shown in Fig.~\ref{fig:04_freq}.

The other suspicious candidate is F1, for the obvious reason of being very close to  $30\cdot f_\text{orb}$ and therefore potentially liable for causing a distortion that survives the folding and binning procedure.
To test our hypothesis, we composed three test cases, omitting either the four suspect modes (\texttt{n03}), or the resonant mode (\texttt{n05}), or both (\texttt{n04}).
As presented in Fig.~\ref{fig:03_res_n}, case \texttt{n03} showed no improvement regarding the model selection, the other two, \texttt{n04} and \texttt{n05} became decisive in the $BIC$ values. 
While the AIC values, which are within the treshold of rejection for the case \texttt{n04},
may suggest that the free axis model has a slightly higher probability of being the real solution, the BIC values and the discovered aligned axis positions indicate that this solution is not stable.

A comparison of Table~\ref{table:1:n00} and Fig.~\ref{fig:03_res_n} reveals that side configurations arise only when the resonant mode F1 is present in the data. We also emphasise that the 'yn' BIC numbers render for all cases the gravity-darkened, free angles models less plausible than the simple, aligned angle models with no gravity darkening.

We note here that $(I_\star, \Omega_\star)$, $(I_\star, 180^\circ \pm \Omega_\star)$ and $(-I_\star, 360^{\circ}-\Omega_\star)$ are all equivalent solutions, because photometry does not distinguish between prograde and retrograde rotations \citep{Barnes2011ApJS..197...10B}. These equivalent configurations clearly show up on Fig.~\ref{fig:03_res_n} as the double peaks in the "violin plots" illustrating the marginal distributions of the angle $\Omega_\star$.

\begin{figure*}
   \resizebox{\hsize}{!}
            {\includegraphics[width=0.7\textwidth]{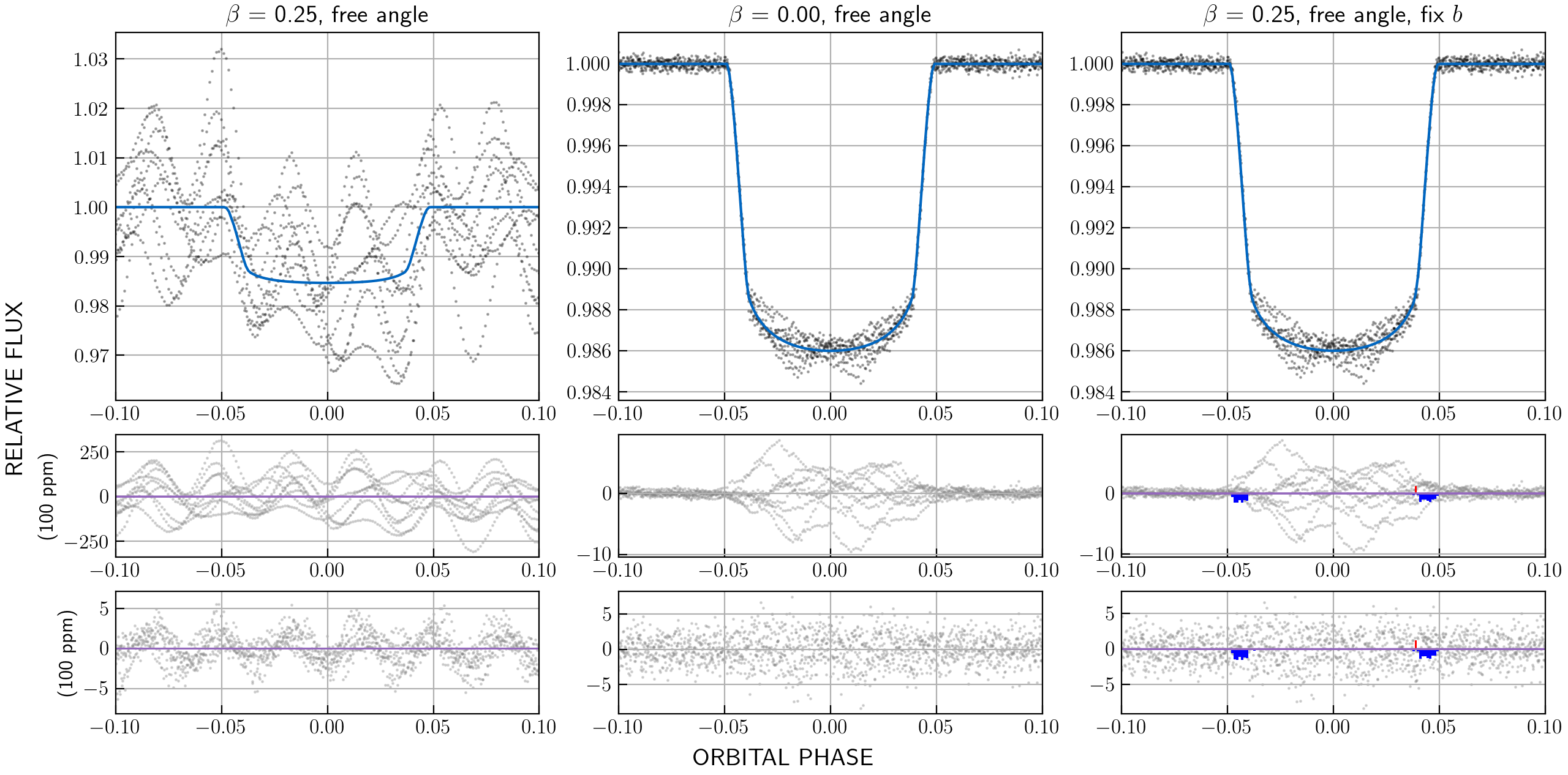}}
      \caption{Summary of the results for the \texttt{pXX} cases. The structure of the figure is similar to Fig.~\ref{fig:01:n02}, 
      except that the left panel 
      corresponds to a data series having all the frequencies, 
      while 
      on the other two columns the 10 detected frequencies were removed as sine model. Therefore the differences between the models are only given for the latter.
              }
         \label{fig:09_res_p}
   \end{figure*}

Regarding the \texttt{pxx} model with enhanced pulsation 
, we found that the limit of capability for the red noise algorithm in \texttt{TLCM} is reached, as the resulting solution exhibits significantly larger uncertainties 
compared to
\texttt{nxx} (see Tab~\ref{table:4:pxx}, leftmost column). The ratio of the uncertainties of \texttt{pxx} to the typical ones from previous runs is around 10; for example, the value and uncertainty of $\Omega_\star$ = $2\substack{+147 \\ -148}$. The lightcurve solution and the original data is shown in the leftmost panels of Fig~\ref{fig:09_res_p}.

\begin{figure*}
   \resizebox{\hsize}{!}
            {\includegraphics[width=0.7\textwidth]{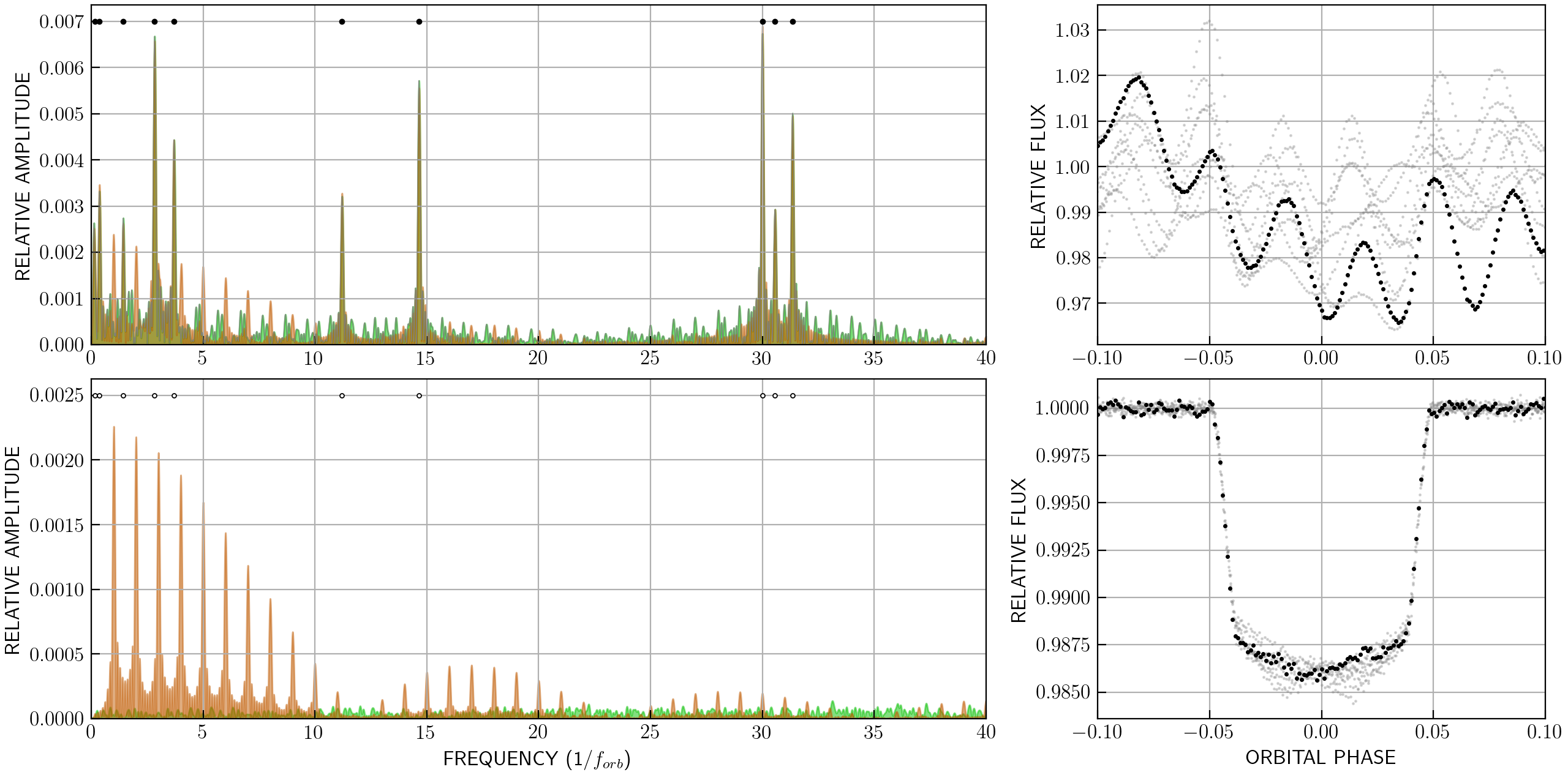}}
      \caption{%
                Illustration of the prewhitening process on the data in the \texttt{pxx} series.
                The top and bottom rows correspond to the data before and after prewhitening, respectively.
                \textbf{Left panels:} the Fourier transform of the full dataset, shown both with the inclusion and exclusion of the transit phases (brown and green colours, respectively). Note that the amplitude of the prewhitened spectrum (bottom left) is shown increased 10 times. The frequencies detected by \texttt{Period04} are shown by filled or empty circles depending on whether  the signal is present or not in the data.
                \textbf{Right panels:} the corresponding folded synthetic dataset (light gray), as well as the data of the first orbital cycle (black).
              }
         \label{fig:10_fou_lc}
   \end{figure*}

Due to this issue, we applied the prewhitening process with \texttt{Period04}, and after identifying the significant frequencies, we subtracted them as a simple harmonic from the dataset, as can be seen in Fig.~\ref{fig:10_fou_lc}. We note that there are some residuals after removal due to the simple model.

As presented in the middle and right columns of Tab.~\ref{table:4:pxx}, the solution obtained using the gravity darkening model exhibits esentially aligned configuration ($I_\star$ = $93\substack{+15 \\ -20}$; $\Omega_\star$~=~$81\substack{+32 \\ -15}$). In any case, the AIC and BIC values also reassure that the simpler model is the correct solution. For completeness, the lightcurve solutions are displayed in the middle and right panels of Fig.~\ref{fig:09_res_p} as well.

\section{Discussion}\label{sec:disc}

Our primary goal was to investigate the possible disturbing effects of pulsations of the host star on the determination of the parameters of a transiting exoplanet. We made the investigation using trial fits on synthetically generated data, contaminated artificially with various types of pulsation signals, with amplitudes exaggerated deliberately to provide a clear study of the effect.

We found that the modulations of nonradial pulsations in the integrated flux caused by the transits are so small that they would be unnoticeable in TESS observations.
Thus the nonradial nature of pulsations can be safely ignored; a radial treatment involving only the diminishing effect of the transits on the amplitudes is sufficient in transit analysis.

The presence of single frequency pulsations did not deter the fitting in any of the investigated cases.
The gravity-darkened, free angles models with $\beta=0.25$ yielded the aligned configuration and essentially the input parameters within errors. The simple models with $\beta=0$ models also restored the same parameters. In addition, the lower AIC and BIC values of the latter indicate that gravity darkening is an unnecessary assumption, not supported by the data.
These criteria become even more decisive when the red noise is considered as a component of the model, confirming that this approach is useful in handling sinusoidal contaminating signals too.

We found, however, a number of multiperiodic, small amplitude pulsation signal combinations which deceived the modelling in the gravity-darkened case (e.g. for \texttt{n02}; see Fig~\ref{fig:01:n02} and Table~\ref{table:1:n02}), and yielded side tilted solutions with $\Omega_\star \sim 0^\circ$. Strangely, intermediate values between 0 and 90 degrees have never occurred. Also, the  stellar inclination essentially stayed close to the plane of the sky, with $I_\star \sim 80-85^\circ$. In this configuration
the $\sim 5-10^\circ$ deviation of the rotation axis from the plane of the sky affects
mainly the profile depth, while the side tilt accounts for small asymmetries in the transit curve. It seems that the pulsation signals mainly affect the latter feature of the transit.

The investigation of the resulted misaligned solutions indicates that they are associated with distortions in the transit light curve caused by the presence of frequencies close to orbital resonance. The other group of of frequencies 
had no such influence. Their indirect resonance is caused by the finite length of the modelled dataset, so the folded and binned light contribution could be different with another number of observations containing more ore fewer transits in question. Therefore this effect depends not only on the frequency of a mode, but on the length of dataset as well. That being said, their removal from the dataset prior to modelling made the $AIC/BIC$-based plausibility evaluation more certain.

Nevertheless, when it comes to model selection, the $\Delta AIC$ and $\Delta BIC$ differences generally point to the model without gravity darkening as the more plausible one.
This is especially true for the 'yn' variant of $\Delta BIC$, because  its lowest values consistently followed the correct solutions in all model scenarios investigated in our work.
This resonates well with the fact that the wavelet method for estimating the correlated (red) noise component is and integral part of \texttt{TLCM}, therefore the 'yn' variants of both $AIC$ and $BIC$ are the proper choice for comparing the different models.

The simulated pulsations should be considered as the lower amplitude regime within typical $\delta$ Sct pulsations. Therefore we also investigated the case of more typical $\delta$ Sct-like pulsations in terms of amplitude. Our findings suggest that the enhanced amplitude reduces the capabilities of \texttt{TLCM} due to the large proportion of red noise in data, making it essential to remove the problematic signals. In our case, 
all frequencies have been cleared as simple harmonics only, which resulted in larger residuals
 during the transit. Despite this fact, the wavelet algorithm of \texttt{TLCM} easily handled the situation, and the aligned axis model was accepted as the correct solution. Note that if other frequencies were present alongside the original 10 frequencies, the additional  residuals could cause problems,
 but addressing that issue beyond the scope of this paper.

Some exoplanetary systems around $\delta$~Scuti type host stars like WASP-33b \citep{2022A&A...660L...2K} and KELT-26b \citep{Rodrigez2020AJ....160..111R} have highly misaligned configurations, some of them close to perpendicular. In case of KELT-26b a spurious frequency close to the 18-th multiple of the orbital frequency was reported, however, it may be unrelated to our findings, given that the misalignment has been confirmed with Doppler tomography. 
The possibility that the pulsations, especially the non-radial ones, could also affect the Doppler tomography data in a similar way cannot be excluded, but such an investigation is clearly beyond the scope of the present paper.

Based on our findings, we propose extreme caution with a simultaneous eclipse or transit modelling that handles a suspected pulsation signal as red noise. Our suggestion is that transit analysis could benefit from cleaning the data of the possible problematic frequencies beforehand, and be prepared to conduct a comparison of the various models using their $BIC$ numbers, even in cases when proper spectroscopy measurements and analysis are available for a better determination of the angles.

\section{Conclusion}\label{sec:conc}

In our work we performed a successful analysis investigating the possible degeneration of pulsations and gravity darkening in the photometric modelling of exoplanet transit light curves.
We constructed dozens of synthetic photometric light curves of increasing complexity with \texttt{pulzem} and \texttt{fitsh} and performed the fits with the \texttt{TLCM} in several predefined runs.
Our results show that the amplitude and phase modulations caused by the transit shadow on the pulsating disc are negligible in the light curve.
We found that the noise handling algorithm of the TLCM generally copes well with datasets containing single mode pulsations, aligned positions were preferred in the false assumption of gravity darkening as well. This conclusion remains generally valid in the presence of multiple frequency signals, with some notable exceptions, however.
According to our findings, frequencies close to resonance with the orbital frequency
may escape the noise handling, cause distortions in the data and thus lead to false results, especially regarding the spin-orbit angles.
This can be avoided by cleaning of the problematic frequencies prior to transit analysis, and comparing the plausibility of tilted models with fixed angles models.
In this respect we found that the most reliable indicator for choosing between the various fitted models is the $BIC$ 'yn' number, which handles the red noise modelling as integral part of the model.
Due to the noise handling wavelet algorithm, there is an upper limit of applicability in the amplitude of the signals. In this case of higher-amplitude pulsations, it is better to clean beforehand the pulsations.
A similar conclusion has been reached by other studies investigating the case of asymmetric transit fits due to the interference of signal coming from stellar activity \citep{Oshaugh2016A&A...593A..25O}.

\begin{acknowledgements}
        We would like to thank the anonymous referee for their suggestions and comments to improve this paper. This work was supported by the PRODEX Experiment Agreement No. 4000137122 between the ELTE E\"otv\"os Lor\'and University and the European Space Agency (ESA-D/SCI-LE-2021-0025). Support of the Lend\"ulet LP2021-9/2021 and the Lend\"ulet LP2018-7/2022 grants of the Hungarian Academy of Science is acknowledged. Project no. C1746651 has been implemented with the support provided by the Ministry of Culture and Innovation of Hungary from the National Research, Development and Innovation Fund, financed under the NVKDP-2021 funding scheme.
\end{acknowledgements}

%
%
\bibliographystyle{aa}
\bibliography{refs}

\begin{appendix}

\section{Tables for multimode, nonradial pulsations}

The tables below present the obtained values from the analysis of \texttt{n00} (Tab.~\ref{table:1:n00}), \texttt{n02} (Tab.~\ref{table:1:n02}), \texttt{pxx} (Tab~\ref{table:4:pxx}) and all the \texttt{nxx} (Tab.~\ref{table:5:nxx}).

\begin{table}
\caption{Summary of fit results for the \texttt{n00} case having 4 frequencies. 
         The conjunction parameter $b$ was fixed during these fits.
         The computed $AIC$ and $BIC$ values, presented at the bottom of the table, clearly favor the more simple model (i.e. without gravity darkening).%
        }\label{table:1:n00}      
\centering          
\begin{tabular}{lrr}
\hline
\hline
{} &         \texttt{n00} fixed angles &              \texttt{n00} model free angles \\
Parameter       &       $\beta = 0.0$                                  &    $\beta = 0.25$                           \\
\hline
$a/R_s$       	     &        $3.6873\substack{+0.0037 \\ -0.0037}$ &	     $3.6682\substack{+0.0093 \\ -0.0050}$ \\[5pt]
$R_p/R_s$	     &     $0.11102\substack{+0.00024 \\ -0.00024}$ &	  $0.10978\substack{+0.00020 \\ -0.00020}$ \\[5pt]
$t_C$         	     &  $0.500047\substack{+0.000084 \\ -0.000084}$ &	  $0.50007\substack{+0.00009 \\ -0.00010}$ \\[5pt]
P         	     &  $0.999995\substack{+0.000017 \\ -0.000017}$ &  $0.999988\substack{+0.000019 \\ -0.000018}$ \\[5pt]
$\sigma_r$ (100 ppm) &  	      $96.3\substack{+1.6 \\ -1.7}$ &		     $96.4\substack{+1.3 \\ -1.3}$ \\[5pt]
$\sigma_w$ (100 ppm) &  	 $1.885\substack{+0.034 \\ -0.035}$ &		$1.887\substack{+0.027 \\ -0.027}$ \\[5pt]
$I_\star$            &  					 90 &			 $94\substack{+12 \\ -15}$ \\[5pt]
$\Omega_\star$       &  					 90 &			$-65\substack{+13 \\ -55}$ \\[5pt]
\hline
$AIC$ yn             &  				    \textbf{-141122} &					   -141081 \\[5pt]
$BIC$ yn             &  				    \textbf{-141080} &					   -141025 \\[5pt]
$AIC$ nn             &  				    -115370 &					   \textbf{-115372} \\[5pt]
$BIC$ nn             &  				    \textbf{-115343} &					   -115330 \\[5pt]
\hline
\hline
\end{tabular}
\end{table}

\begin{table*}
    \caption{%
        Summary of fit results for case \texttt{n02} with all 10 frequencies included.
    }
\label{table:1:n02}      
\centering          
\begin{tabular}{lrrr}
\hline
\hline
{} &  \texttt{n02}; fixed angles \& b fixed &    \texttt{n02}; free angles \& b fitted & \texttt{n02}; free angles \& b fixed \\
Parameter       &  $\beta$ = 0.0 & $\beta$ = 0.25     & $\beta$ = 0.25 \\
\hline
$a/R_s$       	      &       $3.6980\substack{+0.0084 \\ -0.0083}$ &		$3.631\substack{+0.012 \\ -0.012}$ &	    $3.6356\substack{+0.0085 \\ -0.0066}$ \\[5pt]
$R_p/R_s$	      &    $0.11208\substack{+0.00057 \\ -0.00056}$ &	  $0.11081\substack{+0.00045 \\ -0.00044}$ &	 $0.11077\substack{+0.00041 \\ -0.00042}$ \\[5pt]
$t_C$          	      &    $0.49995\substack{+0.00019 \\ -0.00019}$ &	  $0.49996\substack{+0.00011 \\ -0.00011}$ &	 $0.49996\substack{+0.00011 \\ -0.00010}$ \\[5pt]
P         	      & $0.999987\substack{+0.000041 \\ -0.000041}$ &  $0.999982\substack{+0.000028 \\ -0.000025}$ &  $0.999982\substack{+0.000028 \\ -0.000025}$ \\[5pt]
$\sigma_r$ (100 ppm) &  	     $223.9\substack{+3.1 \\ -3.1}$ &		    $223.6\substack{+2.4 \\ -2.3}$ &		   $223.5\substack{+2.3 \\ -2.3}$ \\[5pt]
$\sigma_w$ (100 ppm) &  	 $3.073\substack{+0.066 \\ -0.065}$ &		$3.068\substack{+0.047 \\ -0.048}$ &	       $3.069\substack{+0.049 \\ -0.047}$ \\[5pt]
$b$		      & 					 0. &		$0.001\substack{+0.075 \\ -0.077}$ &					       -1 \\[5pt]
$I_\star $             &					 90 &		     $97.7\substack{+8.7 \\ -9.0}$ &		    $97.7\substack{+8.8 \\ -9.0}$ \\[5pt]
$\Omega_\star$         &					 90 &			  $0\substack{+15 \\ -14}$ &			$-0\substack{+13 \\ -14}$ \\[5pt]
\hline
$AIC$ yn         &   					\textbf{-139526} &				       -139506 &				      -139519 \\[5pt]
$BIC$ yn         &   					\textbf{-139484} &				       -139443 &				      -139463 \\[5pt]
$AIC$ nn         &   					-108523 &				       -108530 &				      \textbf{-108539} \\[5pt]
$BIC$ nn         &   					-108495 &				       -108481 &				      \textbf{-108498} \\[5pt]
\end{tabular}

\end{table*}

\begin{table*}
    \caption{%
        Summary of fit results for case \texttt{pxx} with all 10 frequencies included.
    }
\label{table:4:pxx}      
\centering          
\begin{tabular}{lrrr}
\hline
\hline
{} &  \texttt{pxx}; all freq &    \texttt{pxx} - sin() & \texttt{pxx} - sin() \\
Parameter       &  $\beta$ = 0.25 & $\beta$ = 0.00     & $\beta$ = 0.25 \\
\hline
$a/R_s$       	      &       $3.679\substack{+0.063 \\ -0.073}$ &	  $3.6813\substack{+0.0029 \\ -0.0030}$ &	 $3.6698\substack{+0.0029 \\ -0.0024}$ \\[5pt]
$R_p/R_s$	      &    $0.1165\substack{+0.0036 \\ -0.0036}$ &     $0.11192\substack{+0.00016 \\ -0.00016}$ &     $0.11071\substack{+0.00012 \\ -0.00012}$ \\[5pt]
$b$          	      & 	 $0.11\substack{+0.12 \\ -0.09}$ &					    0. &					    0. \\[5pt]
$t_C$         	      & $0.49988\substack{+0.00083 \\ -0.00080}$ &  $0.500022\substack{+0.000061 \\ -0.000062}$ &  $0.500014\substack{+0.000079 \\ -0.000055}$ \\[5pt]
P                     &  $0.99986\substack{+0.00018 \\ -0.00018}$ &  $0.999998\substack{+0.000013 \\ -0.000012}$ &  $0.999996\substack{+0.000011 \\ -0.000014}$ \\[5pt]
$\sigma_r$ (100 ppm) &  	     $2165\substack{+21 \\ -21}$ &		  $50.4\substack{+2.1 \\ -2.2}$ &		 $50.5\substack{+1.6 \\ -1.6}$ \\[5pt]
$\sigma_w$ (100 ppm) & 	$11.29\substack{+0.14 \\ -0.14}$ &	     $1.966\substack{+0.034 \\ -0.034}$ &	    $1.966\substack{+0.025 \\ -0.025}$ \\[5pt]
$I_\star $             &	       $90\substack{+27 \\ -28}$ &					     90 &		     $93\substack{+15 \\ -20}$ \\[5pt]
$\Omega_\star$         &	      $2\substack{+147 \\ -148}$ &					     90 &		     $81\substack{+32 \\ -15}$ \\[5pt]
\hline
$AIC$ yn         &   				     -135473 &  				    -136278 &					   -136247 \\[5pt]
$BIC$ yn         &   				     -135410 &  				    -136236 &					   -136191 \\[5pt]
$AIC$ nn         &   				      -72379 &  				    -131545 &					   -131495 \\[5pt]
$BIC$ nn         &   				      -72330 &  				    -131517 &					   -131453 \\[5pt]

\end{tabular}

\end{table*}


\longtab[4]{
\begin{landscape}
\footnotesize
\begin{longtable}{l|cccccccc|cccc}
\caption{Summary of fit results for all of the \texttt{nXX} cases having multimode, nonradial pulsations. The conjunction parameter $b$ was fixed during these presented fits.}\\
\label{table:5:nxx}\\
\hline
\hline
Cases; Model &                               $a/R_s$ &                           $R_p/R_s$ &                                        $t_C$ &                                       $P$ &                       $\sigma_{r}$ &                         $\sigma_{w}$ &                         $I_\star$ &                  $\Omega_\star$ &   AIC yn &   BIC yn &   AIC nn &   BIC nn\\
 &  &  &  &  & (100 ppm) & (100 ppm) &  &  &  &  &  & \\[5pt]
\hline
\endfirsthead
\caption{continued}\\
\hline
Cases; Model &                               $a/R_s$ &                           $R_p/R_s$ &                                        $t_C$ &                                       $P$ &                       $\sigma_{r}$ &                         $\sigma_{w}$ &                         $I_\star$ &                  $\Omega_\star$ &   AIC yn &   BIC yn &   AIC nn &   BIC nn\\
 &  &  &  &  & (100 ppm) & (100 ppm) &  &  &  &  &  & \\[5pt]
\hline
\endhead
\hline
\endfoot
\hline
\endlastfoot
\texttt{n00}; ($\beta$ = 0.0)   &  $3.6873\substack{+0.0037 \\ -0.0037}$ &  $0.11102\substack{+0.00024 \\ -0.00024}$ &  $0.500047\substack{+0.000084 \\ -0.000084}$ &  $0.999995\substack{+0.000017 \\ -0.000017}$ &   $96.3\substack{+1.6 \\ -1.7}$ &  $1.885\substack{+0.034 \\ -0.035}$ &				90 &			      90 &  -141122 &  -141080 &  -115370 &  -115343 \\[5pt]
\texttt{n00}; ($\beta$ = 0.25) &  $3.6682\substack{+0.0093 \\ -0.0050}$ &  $0.10978\substack{+0.00020 \\ -0.00020}$ &	  $0.50007\substack{+0.00009 \\ -0.00010}$ &  $0.999988\substack{+0.000019 \\ -0.000018}$ &   $96.4\substack{+1.3 \\ -1.3}$ &  $1.887\substack{+0.027 \\ -0.027}$ &	 $94\substack{+12 \\ -15}$ &  $-65\substack{+13 \\ -55}$ &  -141081 &  -141025 &  -115372 &  -115330 \\[5pt]
\texttt{n01}; ($\beta$ = 0.0)  &  $3.6972\substack{+0.0091 \\ -0.0090}$ &  $0.11227\substack{+0.00058 \\ -0.00057}$ &	  $0.50005\substack{+0.00018 \\ -0.00018}$ &  $0.999965\substack{+0.000037 \\ -0.000038}$ &  $203.0\substack{+3.2 \\ -3.3}$ &  $2.903\substack{+0.068 \\ -0.068}$ &				90 &			      90 &  -139253 &  -139211 &  -110522 &  -110494 \\[5pt]
\texttt{n01}; ($\beta$ = 0.25) &  $3.6363\substack{+0.0089 \\ -0.0080}$ &  $0.11095\substack{+0.00044 \\ -0.00044}$ &	  $0.50001\substack{+0.00012 \\ -0.00011}$ &  $0.999969\substack{+0.000023 \\ -0.000024}$ &  $202.8\substack{+2.3 \\ -2.3}$ &  $2.900\substack{+0.045 \\ -0.047}$ &  $94.6\substack{+7.9 \\ -8.3}$ &	$0\substack{+15 \\ -15}$ &  -139248 &  -139192 &  -110544 &  -110502 \\[5pt]
\texttt{n02}; ($\beta$ = 0.0)  &  $3.6980\substack{+0.0084 \\ -0.0083}$ &  $0.11208\substack{+0.00057 \\ -0.00056}$ &	  $0.49995\substack{+0.00019 \\ -0.00019}$ &  $0.999987\substack{+0.000041 \\ -0.000041}$ &  $223.9\substack{+3.1 \\ -3.1}$ &  $3.073\substack{+0.066 \\ -0.065}$ &				90 &			      90 &  -139526 &  -139484 &  -108523 &  -108495 \\[5pt]
\texttt{n02}; ($\beta$ = 0.25) &  $3.6356\substack{+0.0085 \\ -0.0066}$ &  $0.11077\substack{+0.00041 \\ -0.00042}$ &	  $0.49996\substack{+0.00011 \\ -0.00010}$ &  $0.999982\substack{+0.000028 \\ -0.000025}$ &  $223.5\substack{+2.3 \\ -2.3}$ &  $3.069\substack{+0.049 \\ -0.047}$ &  $97.7\substack{+8.8 \\ -9.0}$ &   $-0\substack{+13 \\ -14}$ &  -139519 &  -139463 &  -108539 &  -108498 \\[5pt]
\texttt{n03}; ($\beta$ = 0.0)  &  $3.6955\substack{+0.0077 \\ -0.0076}$ &  $0.11279\substack{+0.00054 \\ -0.00053}$ &	  $0.49991\substack{+0.00017 \\ -0.00017}$ &  $0.999993\substack{+0.000035 \\ -0.000035}$ &  $215.8\substack{+2.9 \\ -2.9}$ &  $3.010\substack{+0.058 \\ -0.058}$ &				90 &			      90 &  -139274 &  -139232 &  -112326 &  -112298 \\[5pt]
\texttt{n03}; ($\beta$ = 0.25) &  $3.6332\substack{+0.0061 \\ -0.0048}$ &  $0.11147\substack{+0.00041 \\ -0.00041}$ &  $0.499920\substack{+0.000090 \\ -0.000096}$ &  $0.999987\substack{+0.000027 \\ -0.000023}$ &  $215.5\substack{+2.3 \\ -2.2}$ &  $3.008\substack{+0.047 \\ -0.046}$ &  $95.8\substack{+8.3 \\ -8.3}$ &   $-0\substack{+13 \\ -13}$ &  -139242 &  -139186 &  -112348 &  -112306 \\[5pt]
\texttt{n04}; ($\beta$ = 0.0)  &  $3.6871\substack{+0.0082 \\ -0.0079}$ &  $0.11082\substack{+0.00053 \\ -0.00051}$ &     $0.49999\substack{+0.00018 \\ -0.00017}$ &  $1.000009\substack{+0.000037 \\ -0.000037}$ &  $175.6\substack{+2.7 \\ -2.7}$ &  $2.629\substack{+0.064 \\ -0.063}$ &                             90 &                          90 &  -140324 &  -140282 &  -112337 &  -112309 \\[5pt]
\texttt{n04}; ($\beta$ = 0.25) &     $3.665\substack{+0.012 \\ -0.013}$ &  $0.10952\substack{+0.00039 \\ -0.00040}$ &     $0.49997\substack{+0.00016 \\ -0.00013}$ &  $1.000008\substack{+0.000030 \\ -0.000031}$ &  $175.7\substack{+2.1 \\ -2.0}$ &  $2.630\substack{+0.046 \\ -0.046}$ &      $93\substack{+24 \\ -28}$ &  $66\substack{+64 \\ -103}$ &  -140319 &  -140263 &  -112330 &  -112288 \\[5pt]
\texttt{n05}; ($\beta$ = 0.0)  &  $3.6886\substack{+0.0074 \\ -0.0073}$ &  $0.11076\substack{+0.00046 \\ -0.00045}$ &	  $0.50001\substack{+0.00015 \\ -0.00015}$ &  $1.000005\substack{+0.000031 \\ -0.000033}$ &  $179.7\substack{+2.5 \\ -2.5}$ &  $2.653\substack{+0.056 \\ -0.055}$ &				90 &			      90 &  -140549 &  -140507 &  -110299 &  -110271 \\[5pt]
\texttt{n05}; ($\beta$ = 0.25) &     $3.669\substack{+0.016 \\ -0.013}$ &  $0.10949\substack{+0.00046 \\ -0.00045}$ &	  $0.50003\substack{+0.00016 \\ -0.00015}$ &  $0.999996\substack{+0.000033 \\ -0.000031}$ &  $179.7\substack{+2.3 \\ -2.4}$ &  $2.656\substack{+0.050 \\ -0.049}$ &	 $91\substack{+26 \\ -27}$ &   $80\substack{+53 \\ -68}$ &  -140498 &  -140443 &  -110294 &  -110252 \\[5pt]
\end{longtable}
\end{landscape}
}


\end{appendix}

\end{document}